\documentclass{article}
\begin{document}
\title{Analytical calculation of the axis angle $2V$ from extinction measurements on the spindle stage}
\author{F. Dufey}
\maketitle
\begin{abstract}
	A concise derivation of the "Joel equations", which allow for the determination of the axis angle $2V$ from measurements of extinction directions on a spindle stage, is provided starting from the wave-equation. 
	Only analytic methods and no geometric arguments referring to stereographic projections are invoked. 
	For error free data, the resulting equations allow for a closed form solution.  
	If angle data with measurement error are to be used, a maximum likelihood estimation methodology is proposed, which may be solved using e.g.\ iterative reweighting. 
	The method is tested with and compared to  published data. 
\end{abstract}
\section{Introduction}

Since its introduction in the beginning 1950s,  the one-circle or spindle stage goniometer for the determination of the angle $2V$ between the optical axis has become one of the standard methods for this purpose. 
While the determination of the axis angle relied in the first years on rather complicated geometric constructions in the stereographic projection, the technique gained further acceptance when it became possible to solve the underlying equation on a computer \cite{joel1965determination,bloss1973computer,bartelmehs1992excalibr}. 
The derivation of these equations is not easy to comprehend today, as it involves a mix of geometric and analytic constructs, like the equivibration curve,  and is scattered several articles \cite{joel1957theextinction, garaycochea1964determination, joel1963determination,joel1965determination}. 
It is one goal of this notice to give a concise derivation of these equations in a purely analytical fashion starting from the Fresnel equation. 
The second result to be presented is that these equations principally allow for a closed form solution for the components of the inverse dielectric tensor $\Phi$, and hence also for $2V$, when error free extinction positions are available and to derive a maximum likelihood estimator for the components of the inverse dielectric tensor, when data prone to statistical error are used. 

\section{Derivation of a formula for the angle $2V$}

The tensor $\Phi$ is the inverse of the optical dielectric tensor,
\[ 
	\Phi=\epsilon^{-1}.
\]
It can be decomposed as

\[
	\Phi=A_2 \mathbf{1} -\frac{1}{2} (A_1-A_3) \varphi,
\]
Here, $A_i=1/\epsilon_i=1/n_i^2$ ($i \in \{1,2,3\}$) where $\epsilon_i$ are the principal components of the dielectric  tensor and $n_i$ the main indices of refraction. 
The $A_i$ are ordered as $A_1>A_2>A_3$ (uniaxial crystals are not considered here).
The tensor $\varphi=a_1a_2^T+a_2a_1^T$ is made up from the two vectors $a_1$ and $a_2$ of unit length which are perpendicular to the two circular sections of the indicatrix and parallel to the optical axes. 
We are interested in calculating  the tensor $\Phi$ and the angle $2V$ between the vectors $a_1$ and $a_2$, but will not use these vectors any further in the course of the following calculations.

According to \cite{landau2013electrodynamics}, eq.\ 21, the directions of extinction coincide with the dielectric displacement vector $D$, which fulfills the wave equation
\begin{equation}
	\label{eq:wave}
	(\mathbf{1}-\hat{n}\hat{n}^T)\Phi D =\frac{1}{n^2}D,
\end{equation}
where $\hat{n}=k/|k|$ is the unit vector parallel to the wavevector $k$  of the wave. 
The resulting equation for the refractive index $n$ is Fresnel's equation, but we are more interested in the measured direction of 
$D$, which is always perpendicular to $\hat{n}$.

So if we hold $\hat{n}$ and $D$ fixed and choose the coordinate axes as $e_y=\hat{n}$ and $D || e_z$, (the direction $e_x$ is then parallel to the second possible polarisation.) the equation for the $z$-component is trivially fulfilled and the equation for the $y$-component can always be fulfilled with a suitable choice of $n$, leaving the only restriction, that the left hand side does not introduce a y-component for $D$, or,
\begin{equation}
	\label{eq:Phixy}
	\Phi_{xy}=e_y^T\Phi e_x=0.
\end{equation}
Now for any two vectors $q$ and $q'$ which fulfill $(q+q')|| e_x$ and $(q-q')||e_y$, we get
\[
	(q+q')\Phi(q-q')=0,
\]
which are the equations given by Joel. 

Before solving eq.\ \ref{eq:Phixy}, we note that the tensor $\Phi$ is a symmetric tensor which can be parametrized by 6 constants. 
It is clear that the direction of $D$ cannot depend on the isotropic part of $\Phi$, 
\[
	\Phi_\mathrm{iso}= \mathbf{1}\;(\Phi_{xx}+\Phi_{yy}+\Phi_{zz})/3
\]
 and is also invariant under a rescaling $\Phi \rightarrow c\Phi$ with an arbitrary constant c.
Therefore, we can hope at best to recover from measurements on the spindle stage the anisotropic part 
\[
	\Phi_\mathrm{aniso}=\Phi -\Phi_\mathrm{iso}
\] 
which depends on 5 parameters of which one may be chosen at will to fix the scaling. 
An explicit expression for $\Phi_\mathrm{aniso}$ is 
\[
    \Phi_\mathrm{aniso}=	\left(\begin{array}{ccc}
			d_{x^2-y^2}-d_{z^2} & d_{xy} & d_{xz}\\
			d_{xy} &-d_{x^2-y^2}-d_{z^2}&d_{yz}\\
			d_{xz}&d_{yz}& 2d_{z^2}
	\end{array}\right),
\]
with 

\begin{eqnarray*}
	d_{z^2}&=&(2\Phi_{zz}-\Phi_{xx}-\Phi_{yy})/6,\\
	d_{x^2-y^2}&=&(\Phi_{xx}-\Phi_{yy})/2,\\
	d_{xy}&=&\Phi_{xy},\\
	d_{xz}&=&\Phi_{xz},\\
	d_{yz}&=&\Phi_{yz}.
\end{eqnarray*}

Various orientations parameterized by the spindle angle $S$ have to be analysed 
which results in the substitution. 
\[
	\Phi\rightarrow \tilde{\Phi}(S, E_S)=R_y(E_S)R_z(S)\;\Phi \;R^T_z(S)R^T_y(E_S).
\]
Here, $R_y$ and $R_z$ are the usual rotation matrices around the $y$- and $z$-axes, respectively,
\[
R_y=\left( \begin{array}{ccc} \cos E_S &0 & -\sin E_S \\ 0 & 1 &0 \\ \sin E_S  & 0 & \cos E_S \end{array} \right), 
\] 
and 
\[
R_z=\left( \begin{array}{ccc}  \cos S & \sin S & 0 \\ -\sin S & \cos S & 0 \\ 0 & 0 & 1 \end{array} \right). 
\] 
Specifically, $z$ is chosen to be the spindle axis and $y$ the direction perpendicular to the rotation desk of the microscope with $S$ and $E_S$ being the respective rotation angles.

Substitution into eq.\ \ref{eq:Phixy} yields an implicit equation for the equivibration curve $E_S(S)$, 

\begin{eqnarray*}
	-3d_{z^2}\tan 2E_S+2d_{xz}\cos S+ 2d_{yz}\sin S+\\+ d_{x^2-y^2}\cos 2 S\tan 2E_S+d_{xy}\sin 2S\tan 2E_S=0.
\end{eqnarray*}

Hence one of the  d-terms may be fixed advantageously, e.g.\ $d_{z^2}=1$. 
Then, if measurements are made for 4 angles $E$, this linear system for the other $d_i$ can be solved exactly. 
However, if more than 4 angles are measured, then some regression technique has to be employed.
The programs available \cite{bloss1973computer,bartelmehs1992excalibr} minimize an ad hoc sum of squared Joel equations. 
This arbitrarily assigns weights ($=1$) to all observations. 
To put the regression on a statistical basis, it seems more appropriate to start from a reasonable error model for the dependent variables $E_S$. 
With the expectation value of the angle $E_S$ being $1/2\,\arctan{V/W}$ where $V=2d_{xz}\cos S+ 2d_{yz}\sin S$ and $W=3d_{z^2}-d_{x^2-y^2}\cos 2 S-d_{xy}\sin 2S$, 
we assume the measurement error of the angles to follow a von Mises distribution \cite{evans2000statistical}, 
\[
	f(2E_S)= \exp(\pm\kappa \cos(2E_S-\arctan(V/W)))/2\pi I_0(\kappa)
\]
where $\kappa$ is a dispersion parameter and $I_0$ is a modified Bessel function.
The sign of the exponent depends on which of the two solutions $E_S$ is chosen with points on the same extinction curve giving rise to the same sign. 
The parameter $\kappa$ may be estimated as the empirical variance of the observed angles from the calculated values. 
The likelihood to be maximized with respect to the parameters $d_j$, $j\in \{x^2-y^2,xy, xz, yz\}$ given measured values $E_{Si}$ from one of the two extinction curves with $i\in [1,n]$  is then
\[
	\label{eq:L}
	L(\{d_j\})=\pm\kappa\sum_{i=1}^N \frac{W_i\cos 2E_{Si} +V_i \sin 2E_{Si} }{\sqrt{V_i^2+W_i^2}} 
\]
The optimum is attained when the four equations 
\[
	0=\frac{\partial L}{\partial d_j}=\pm \kappa\sum_i \left(W_i \sin 2E_{Si}-V_i \cos{2E_{Si}}\right) (V_i^2+W_i^2)^{-3/2} \frac{\partial V_iW_i}{\partial d_j}
\]
are fulfilled, which can be seen to be weighed sums of the original Joel equations. 
Hence, iterative reweighting seems to be a promising and numerically stable option to obtain maximum likelihood estimates of the parameters $d_j$. 
The second derivatives of $L$ with respect to the $d_j$ yield the information matrix from which estimates of the variances of the $d_j$ may be obtained. 

The eigenvectors of $\Phi_\mathrm{aniso}$, constructed from the estimates $d_j$,  are the directions of the main axes of the indicatrix, 
corresponding to the eigenvalues $\mu_\mathrm{h}$, $\mu_\mathrm{m}$ and $\mu_\mathrm{l}$, ordered by decreasing value. 
From the latter, the axis angle $2V$ can be calculated \cite{garaycochea1964determination},
\[
	\tan^2{V}=\frac{\mu_\mathrm{h}-\mu_\mathrm{m}}{\mu_\mathrm{m}-\mu_\mathrm{l}}.
\]
It is obvious that this result is not influenced by neither a  constant scaling nor a shift of all eigenvalues. 

\section{Comparison with published data}

To exemplify the feasibility of the method, angles $2V$ calculated from the measured angles $E_S$ taken from  the  articles \cite{bloss1973computer, bartelmehs1992excalibr, gunter2004results}
are being compared to values $2V$ reported there: 

\cite{bloss1973computer}, Fig.\ 5: 64.09$^\circ$ (63.66$^\circ$)

\cite{bartelmehs1992excalibr}, Fig.\ 2: 76.88$^\circ$ (77.57$^\circ$)

\cite {gunter2004results}, Fig.\ 4:   48.98$^\circ$ (49.36$^\circ$)

The numerical optimization of the likelihood eq.\ \ref{eq:L} was performed using the optmodel procedure of  \cite{SAS}.  

Expected values $E_S$ calculated with these parameters compare very well with the reported ones with the mean squared deviations from the measured $E_S$ being even slightly smaller.

\bibliography{extinction}{}
\bibliographystyle{apalike}
\end{document}